\documentclass[conference]{IEEEtran}
\IEEEoverridecommandlockouts
\usepackage{cite}
\usepackage{amsmath,amssymb,amsfonts}
\usepackage{algorithmic}
\usepackage{graphicx}
\usepackage{textcomp}
\usepackage{xcolor}
\def\BibTeX{{\rm B\kern-.05em{\sc i\kern-.025em b}\kern-.08em
    T\kern-.1667em\lower.7ex\hbox{E}\kern-.125emX}}
    
\usepackage{color}
\usepackage{braket}
\usepackage{url}
\newcommand{\vect}[1]{\boldsymbol{#1}}
\usepackage{mathtools}
\usepackage{qcircuit}
\usepackage{todonotes}

\DeclarePairedDelimiter\ceil{\lceil}{\rceil}

\begin{document}

\title{Trainable Discrete Feature Embeddings for\\ Variational Quantum Classifier
}

\makeatletter
\newcommand{\linebreakand}{%
  \end{@IEEEauthorhalign}
  \hfill\mbox{}\par
  \mbox{}\hfill\begin{@IEEEauthorhalign}
}
\makeatother

\author{\IEEEauthorblockN{Napat Thumwanit}
\IEEEauthorblockA{\textit{Dept. of Computer Science} \\
\textit{The University of Tokyo}\\
Bunkyo-ku, Tokyo, Japan \\
thumwanit-napat@g.ecc.u-tokyo.ac.jp}
\\
\IEEEauthorblockN{Hiroshi Yano}
\IEEEauthorblockA{\textit{Dept. of Applied Physics and Physico-Informatics} \\
\textit{Keio University}\\
Kohoku-ku, Yokohama, Japan \\
hiroshi.yano.gongon@keio.jp}
\and
\IEEEauthorblockN{Chayaphol Lortaraprasert}
\IEEEauthorblockA{\textit{Dept. of Mechanical Engineering} \\
\textit{The University of Tokyo}\\
Bunkyo-ku, Tokyo, Japan \\
lortaraprasert@g.ecc.u-tokyo.ac.jp}
\\
\IEEEauthorblockN{Rudy Raymond}
\IEEEauthorblockA{\textit{IBM Quantum} \\
\textit{IBM Japan}\\
Chuo-ku, Tokyo, Japan\\
rudyhar@jp.ibm.com}
}

\maketitle

\begin{abstract}
Quantum classifiers provide sophisticated embeddings of input data in Hilbert space promising quantum advantage. 
The advantage stems from quantum feature maps encoding the inputs into quantum states with variational quantum circuits. 
A recent work shows how to map discrete features with fewer quantum bits using \textit{Quantum Random Access Coding} (QRAC), an important primitive to encode binary strings into quantum states. 
We propose a new method to embed discrete features with trainable quantum circuits by combining QRAC and a recently proposed strategy for training quantum feature map called \textit{quantum metric learning}. 
We show that the proposed trainable embedding requires not only as few qubits as QRAC but  also overcomes the limitations of QRAC to classify inputs whose classes are based on hard Boolean functions. 
We numerically demonstrate its use in variational quantum classifiers to achieve better performances in classifying real-world datasets, and thus its possibility to leverage near-term quantum computers for quantum machine learning.  
\end{abstract}


\section{Introduction}
Quantum computers can execute algorithms based on the principles of quantum mechanics and can provide quantum speedups over classical computers in wide-range applications, including machine learning~\cite{Biamonte2017}. There are many proposed quantum algorithms for machine learning that are proven superior to their classical counterparts, such as, solving linear equations~\cite{HHL2009} and principal component analysis~\cite{Lloyd_2014}, all of which require perfect quantum computers. Recent noisy quantum devices~\cite{Preskill2018} are built towards the goal of realizing such perfect quantum computers with the progress comparable to that of the Moore's Law~\cite{jurcevic2020demonstration}. Nevertheless, although limited in the number and quality of their quantum bits (or, qubits), the recent hardware progress has sparked active studies in the search of quantum methods that are applicable to near-term quantum devices~\cite{mcclean2016theory,kandala2017hardware,Havlcek2019SupervisedLW}.

Variational Quantum Classifiers (VQC) is a potential machine learning method on near-term quantum devices for classifying classical data. VQC is built from variational quantum circuits whose parameters are tuned by coordinating the noisy quantum devices and perfect classical computers. It finds a quantum feature map embedding classical data into a high-dimensional Hilbert space, and a measurement set to perform the classification. The most popular type of VQCs fixes the quantum feature map for each classical data, and trains a measurement for the classification. This is akin to quantization of the classical kernel methods, such as the polynomial support vector machine (SVM)~\cite{goldberg-elhadad-2008-splitsvm}. There have been various proposals of quantum feature maps embedding real-valued data with near-term quantum devices~\cite{schuld2018circuitcentric,Schuld_2019,Mitarai_2018,Havlcek2019SupervisedLW}. A recent work~\cite{Yano2020EfficientDF} shows how to use QRAC in VQCs for mapping binary strings into the Hilbert space. The resulting VQCs are shown to require fewer qubits while retaining high classification accuracy.

QRAC encodes $m$ classical bits with $n$ qubits so that any $1$-out-of-$m$ bits can be recovered with probability $p > 1/2$ \cite{ANTV02}. 
It is often denoted as $(m, n, p)$-QRAC). When $p = 1$, the Holevo bound~\cite{Holevo1973} implies that $n \ge m$ meaning no quantum advantage. But when $1/2 < p < 1$, QRAC can achieve provable quantum advantage, e.g., in quantum communication~\cite{HINRY07,quadeer2017strong}. In particular, when $p$ is enough to be bigger than $1/2$, the number of qubits of QRAC is shown exactly half of the bits for classical encoding~\cite{INRY07}. 

The advantage of QRACs stems from the facts that ($i$) the $n$-qubit quantum states are spanned by the $2^n$-dimensional Hilbert space whose real dimension is $2^{2n}-1$, and ($ii$) bits of length up to $2^{2n}-1$ can be mapped into quantum states distributed in the Hilbert space. The QRACs for mapping discrete features in VQCs in~\cite{Yano2020EfficientDF} are mainly with the well-known construction of $(m, n, p)$-QRACs for $n = 1, 2$. Main challenges in using QRACs for quantum feature map are their limited known construction, namely, no optimal QRAC is known for three or more qubits, and the difficulties to map bitstrings to quantum states that guarantees maximally separating embeddings. The latter one is serious because the fixed mapping by QRACs can result in an arrangement of quantum states which is impossible to find its boundary decision without adding more resources such as latent qubits or copies of encoding states. This can mitigate the benefits of QRACs to map discrete features with fewer qubits. 

To solve the aforementioned challenges, we extend the use of QRACs to embed discrete features with the new strategy of \textit{quantum metric learning}~\cite{Lloyd2020QuantumEF} to obtain trainable embeddings. The trainable discrete feature embeddings can find well-separated quantum states in the Hilbert space by taking into account the property of QRACs to use a set of diverse quantum states for encoding discrete features. By numerical experiments, we show the limitations 
of QRACs on classifying inputs whose decision depends on hard Boolean functions called \textit{parity}, and show how trainable embeddings can overcome them. We also provide evidences the trainable embeddings can maintain the advantage of QRACs to achieve high classification accuracy with fewer qubits on various real-world datasets such as the breast cancer dataset \cite{Dua:2019}, the Titanic survival datasets \cite{titanic}, and the simplified version of MNIST \cite{lecun2010mnist}. The source code of our implementation of trainable embedding is available at \url{https://github.com/barnrang/Trainable-Embedding-QML}. We believe the trainable embeddings provide new insights on how to levarage near-term quantum devices for machine learning.

\section{Related Work}\label{sec:related}
Dealing with discrete features, representing qualitative or categorical properties, is 
important because many real-world datasets contain such features, such as, gender, race, age group, etc.
We propose a new framework to embed the discrete features in VQC for quantum machine learning. 
Classical machine learning models such as decision trees~\cite{cart84} can deal directly with discrete features, but 
models such as neural-network often rely on transforming them into continuous features~\cite{Hancock2020}. 
Standard transformations include entity embeddings~\cite{GuoB16} and distributed representations~\cite{hinton1986learning}. Most successful transformation often rely on combination of transformation methods, e.g., \textit{word2vec}~\cite{Mikolov2013EfficientEO} obtains continuous representation of categorical features from the use of one-hot encoding and neural networks. 

Quantum-enhanced machine learning techniques such as VQCs~\cite{schuld2018circuitcentric,Havlcek2019SupervisedLW} are quite similar to 
neural networks and their framework are often developed by assuming appropriate representation of datasets with real-valued and continuous features. VQCs consists of a series of quantum operations acting on qubits represented as a quantum circuit. The 
circuit can be divided into two: the part for mapping classical data into 
quantum states, and the part for measuring the quantum states to obtain the classification classes. The mapping part is usually fixed and there are many potential methods that maybe difficult to realize with classical computers. Recently, \cite{Lloyd2020QuantumEF} proposes trainable quantum feature mapping (or, embedding) called quantum metric learning, akin to the classical metric learning~\cite{CHL95}, so that the measurement part can be simplified. The measurement part is usually a parameterized circuits whose optimal parameters are computed with variational methods similar to those of neural networks. A recent work~\cite{Yano2020EfficientDF} shows how to directly deal with discrete features in VQCs by using QRACs that can reduce computational resources of VQCs. 

Originally formulated in the communication setting~\cite{Wiesner1983}, QRACs have been extensively used in the theory of quantum computations, such as, quantum state learning \cite{Aaronson07}. Some QRACs also offer cryptographic properties known as {\textit{parity obliviousness}} \cite{CKKS16}, which can be useful for private information retrieval. QRACs can encode $n$-bit binary strings using $\log{n}/{2}$ qubits, which is half of the bits used in classical random access codes (RACs)~\cite{INRY07}. 
QRACs with 
one and two qubits were shown in~\cite{ANTV02,L17,IR2018}. One qubit can encode at most three bits, and two qubits at most fifteen bits, and so on. \cite{Yano2020EfficientDF} shows how to use them for healthcare datasets, but there are still many obstacles to deal with complicated decision functions. Such QRACs may be used as a layer in a quantum neural network (QNN)~\cite{Farhi2018ClassificationWQ,Schuld2014TheQF}.

There are variants of QRACs using shared entanglement and classical randomness~\cite{ALMO08}, 
that enable encoding any number of bits into a single qubit. Like the QRACs and our proposed embedding, those variants rely on finding a set of quantum states that are as diverse as possible within the Hilbert space of the underlying qubits.

\section{Methods}\label{sec:proposed}
Our method deals with embedding discrete features of inputs into the Hilbert space and finding the boundary decision using VQCs. We briefly describe the quantum circuits, 
VQCs and their embeddings before showing the proposed method of Trainable Embeddings (TEs).  

\subsection{Preliminaries}

\subsubsection{Quantum Circuit (QC)} QC is popular to model quantum computation. 
It consists of sequences of quantum gates that transform qubits from initial quantum states (usually the all-zero state) into final quantum states that can be measured to obtain the computational result (such as, the classification). An $n$-qubit (pure) quantum state $\ket{\psi}$ is a linear superposition of all possible $n$-bit classical states $\ket{i}$ for $i = 0, \ldots, 2^n-1$. Namely, $\ket{\psi} = \sum_{i} \alpha_i \ket{i}$, where $\alpha_i \in \mathbb{C}$ is the probability amplitude whose squared magnitude is the probability of measuring $\ket{i}$, and thus $\sum_{i} \|\alpha_i\|^2 = 1$. The $n$-qubit quantum state is thus a complex vector lies in a $2^n$-dimensional Hilbert space. Quantum gates are unitary operations that can be represented by a series of $2^n \times 2^n$ unitary matrices $\{\mathbf{U}_i\}$ rotating $2^n$-dimensional complex vectors. All unitary operators can be realized with a small set of universal gates that consists of arbitrary one-qubit rotation and two-qubit entangling gates, such as the controlled-NOT and the ZZ-interaction gate. See~\cite{Nielsen2000} for more details on QC. The near-term quantum computers are those with dozens to hundreds of qubits, and can only run few quantum gates. Therefore, all quantum machine learning leveraging near-term devices must be designed with shallow QCs in mind.

\subsubsection{Variational Quantum Classifiers (VQC)} %
VQC is a classifier using variational quantum circuit that can be trained from labelled inputs, $S =\{(\vect{x}_i,y_i)\}_i$ for $\vect{x}_i \in \mathbb{R}^d$ and $y_i \in \{0,1\}$. As in Fig.~\ref{fig:vqc}, it consists of two parts: the first implementing feature map of input data, namely, $\vect{x}: \mathbf{\Phi(\vect{x})} \in \mathbb{R}^k$ for $k \ge d$, and the second implementing measurement to determine the classification. Its principle is similar to support vector machine (SVM)~\cite{Cortes95support-vectornetworks} for finding the best hyperplane $(\boldsymbol{w}, b)$ that classifies 
the embedded data by the linear boundary function $f(\boldsymbol{x}) = \vect{w}^T \mathbf{\Phi}(\boldsymbol{x}) + b$.

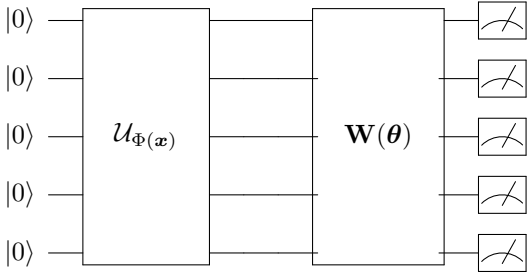
\begin{figure}[t!]
    \centering
    \mbox{
    \Qcircuit @C=1.3em @R=0.8em {
         \lstick{\ket{0}}  & \multigate{4}{~~\mathcal{U}_{\Phi(\boldsymbol{x})}~~}      & \qw    & \qw & \multigate{4}{~~\mathbf{W}(\boldsymbol{\theta})~~}  & \meter &   &\\
         \lstick{\ket{0}}  & \ghost{~~\mathcal{U}_{\Phi(\boldsymbol{x})}~~}             & \qw    & \qw & \ghost{~~\mathbf{W}(\boldsymbol{\theta}~~}          & \meter &   & \\
         \lstick{\ket{0}}  & \ghost{~~\mathcal{U}_{\Phi(\boldsymbol{x})}~~}             & \qw    & \qw & \ghost{~~\mathbf{W}(\boldsymbol{\theta}~~}          & \meter &  &  \\
         \lstick{\ket{0}}  & \ghost{~~\mathcal{U}_{\Phi(\boldsymbol{x})}~~}             & \qw    & \qw & \ghost{~~\mathbf{W}(\boldsymbol{\theta}~~}          & \meter &  &\\
         \lstick{\ket{0}}  & \ghost{~~\mathcal{U}_{\Phi(\boldsymbol{x})}~~}             & \qw    & \qw & \ghost{~~\mathbf{W}(\boldsymbol{\theta}~~}          & \meter &  &\\
    }
    }
        \caption{A VQC that consists two subcircuits: $\mathcal{U}_{\Phi(\boldsymbol{x})}$ for~mapping features and $\mathbf{W}(\boldsymbol{\theta})$ for boundary decision.}
        \label{fig:vqc}
\end{figure}

At VQC, the data $\boldsymbol{x} \in \mathbb{R}^d$ is mapped to (pure) quantum state 
by the feature map circuit $\mathbf{U}_{\Phi(\boldsymbol{x})}$ that realizes $\Phi(\boldsymbol{x})$. 
That is, conditioned on the data $\boldsymbol{x}$ we apply the circuit $\mathbf{U}_{\Phi(\boldsymbol{x})}$ 
to the $n$-qubit all-zero state $\left|0_n\right>$ to obtain the quantum state $\left|\Phi(\boldsymbol{x})\right>$. 
A short-depth quantum circuit $\mathbf{W}(\boldsymbol{\theta})$ is then applied to the quantum state, 
where $\boldsymbol{\theta}$ is the hyperparameter set of the quantum circuit that can be learned from the training data. 
Finding the circuit $\mathbf{W}(\boldsymbol{\theta})$ is akin to finding the separating hyperplane $(\boldsymbol{w},b)$ 
in the Soft-SVM, with the promise of quantum advantage due to the difficulty for 
classical procedures to realize the feature map $\Phi(\boldsymbol{x})$.

Learning the best $\boldsymbol{\theta}$ can be obtained by minimizing the empirical risk $R(\boldsymbol{\theta})$ 
with regards to the training data $\mathit{S} = \left\{ (\boldsymbol{x}_i, y_i)\right\}$. The risk is then approximated with a continuous function, as detailed in \cite{Havlcek2019SupervisedLW}, to enable variational methods for tuning $\boldsymbol{\theta}$ to minimize the approximated cost function.

The binary classification with VQC now follows from first fixing the feature maps on inputs and then training the VQC to learn the best $\boldsymbol{\theta}^*$ to obtain $(\boldsymbol{w}(\boldsymbol{\theta}^*), b^*)$. 
The classification against unseen data $\boldsymbol{x}$ is then performed according to the classifier 
function $f(\boldsymbol{x})$ with $(\boldsymbol{w}(\boldsymbol{\theta}^*), b^*)$. 
Both training and classification need to be repeated for multiple times (or, shots) 
due to the probabilistic nature of quantum computation.

\subsubsection{Quantum Neural Networks (QNN)} 
We use the QNN proposed by \cite{Farhi2018ClassificationWQ} for binary classification with binary inputs $\vect{x} \in \{0,1\}^n$. It is a special case of VQC in Fig.~\ref{fig:vqc}. At QNN two quantum registers are used; the first to represent the $n$-bit input and the second for its label, and thus requires at least $n+1$ qubits (omitting the working qubits used in the middle of the computation). QNN is VQC with simple $n$-qubit feature map $\Phi(\vect{x})$ to create the quantum state representing computational basis $\ket{\vect{x}}\ket{0} = \mathbf{U}_{\Phi(\vect{x})} \ket{0}^{\otimes n}\ket{0}$. This can realized with the $\mathbf{X}$ gate at the $i$-th qubit for $x_i = 1$.   

At QNN $\mathbf{W}(\boldsymbol{\theta})$ is a sequence of two-qubit unitaries $\mathbf{U}_1(\theta_1),\mathbf{U}_2(\theta_2)\ldots ,\mathbf{U}_l(\theta_l)$. The optimization target is to find the set of parameters $\boldsymbol{\theta}=\{\theta_1,\theta_2,\ldots,\theta_l\}$
that minimizes the training loss of dataset $\mathcal{D}$ such as binary cross-entropy or hinge loss. The evolution of the state vector due to $\mathbf{W}(\boldsymbol{\theta})$ is $\ket{\psi}=\mathbf{W}(\boldsymbol{\theta})\ket{\vect{x},0} = \sum_{i} \left(\alpha_{i0} \ket{i,0} + \alpha_{i1} \ket{i,1}\right)$, 
where the classification $l$ is obtained by measuring the second register, i.e., $l = 0$ iff. $\sum_{i0} |\alpha_{i0}|^2 \ge \sum_{i1} |\alpha_{i1}|^2$. \cite{Farhi2018ClassificationWQ} showed how to perform supervised learning on QNN for any $n$-bit Boolean functions, and thus for binary classification of discrete inputs. In particular, numerical experiments showed that QNN could learn the parity function of a subset of the input bits with samples much less than $2^n$ even under some noisy labels. However, \cite{Farhi2018ClassificationWQ} pointed out that the learning became impossible as the subsets became larger. This motivates our choice of the parity functions in Experiments. 

\subsubsection{Discrete Feature Encoding with QRAC}
Discrete features can be trivially mapped into binary strings and be used as inputs to VQCs. Here, we consider binary representation. For example, a categorical feature with at most $c$ different categories can be represented with $\ceil{\log{c}}$ bits and thus requires the same amount of qubits at VQCs. In practice, if there are $k$ types of discrete features, and each of the features has $c_i$ categories, then the number of bits representing all the features is at most $S = \sum_{i=1}^{k}\ceil{\log_2(c_i)}$.

\cite{Yano2020EfficientDF} showed how to use QRAC to reduce the number of qubits representing the categorical features. In particular, it is known how to encode bitstrings of length $2$ and $3$ by $(2,1)$-QRAC and $(3,1)$-QRAC (we omit the success probabilities $p$) by mapping each bitstring into a one-qubit quantum state. The left figure in Fig.~\ref{fig:3bits_parity} depicts the states of $(3,1)$-QRACs that are placed at the corners of a cube in the Bloch sphere. Due to this property of evenly placing quantum states, QRAC ensures that each quantum encoding of a string can be distinguished from other encodings. This can be advantageous for classification with few qubits. By using $(m,n)$-QRAC the number of qubits required for encoding $S$ bits is $n\times\ceil{S/m}$. 

The discrete feature encoding by QRAC is easily combined with continuous features. For discrete  feature $\vect{x}_d$ and continuous feature $\vect{x}_c$, $\vect{x}_d$ can be embedded using $(m,n)$-QRAC and $\vect{x}_c$ using some continuous feature map such as the ZZ feature map~\footnote{The details of the ZZ feature map can be found at \url{https://qiskit.org/documentation/stubs/qiskit.circuit.library.ZZFeatureMap.html}}. 
We can also apply QRAC in the convolutional manner that we call \textit{C-QRAC} (for \textit{Convolutional}-QRAC). For example, with regard to binary strings $\vect{b} = b_1b_2\ldots b_t$, instead of encoding it sequentially, 
the convolutional manner can encode its overlapped substrings. The convolutional preprocessing is effective for image datasets, and therefore we believe so is the C-QRAC for MNIST datasets.

\subsection{Trainable Embedding (TE)}

\begin{figure*}[tb]%
\begin{minipage}[t]{0.3\textwidth}
\centering
    \includegraphics[width=\textwidth]{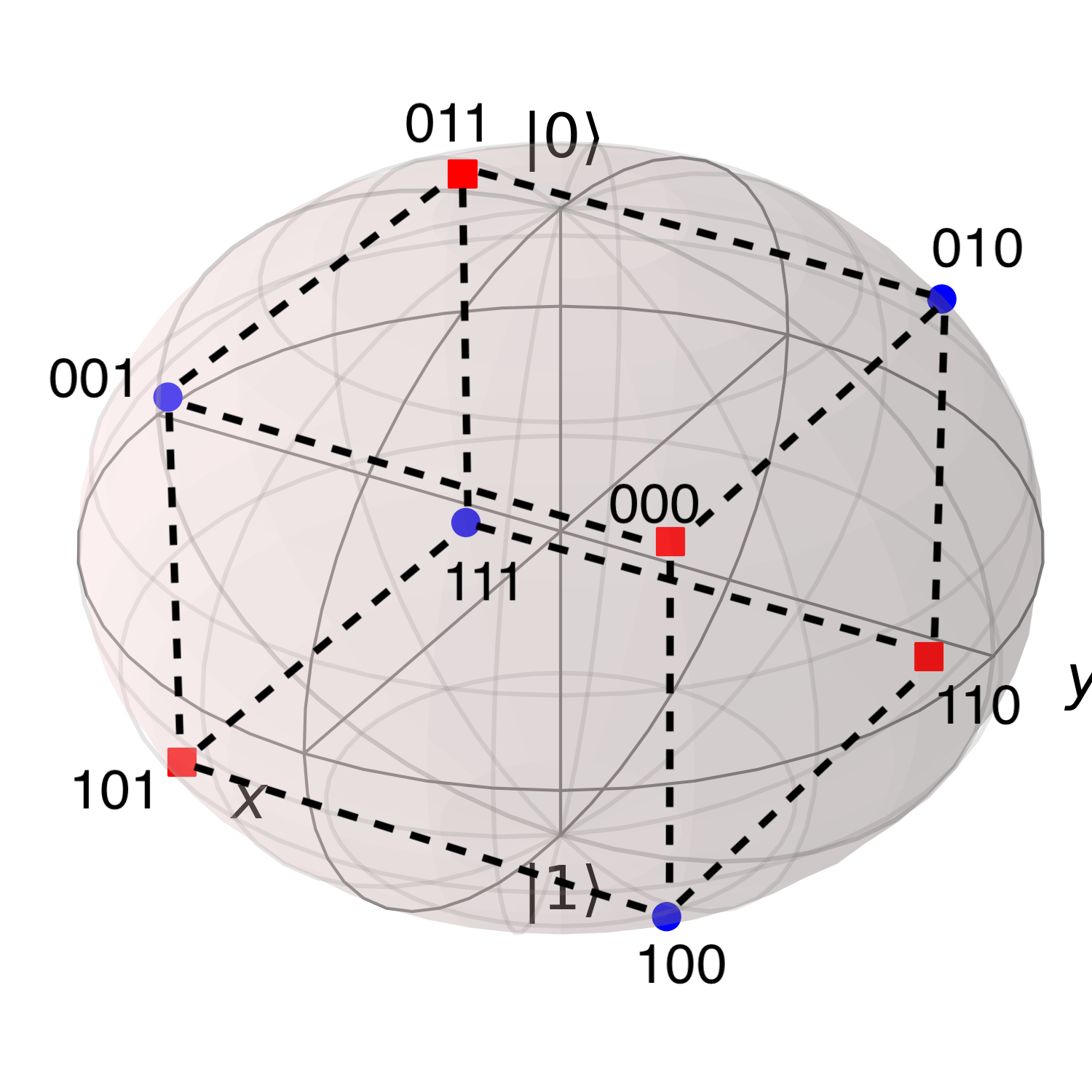}
\end{minipage}%
\begin{minipage}[t]{0.3\textwidth}
\centering
    \includegraphics[width=\textwidth]{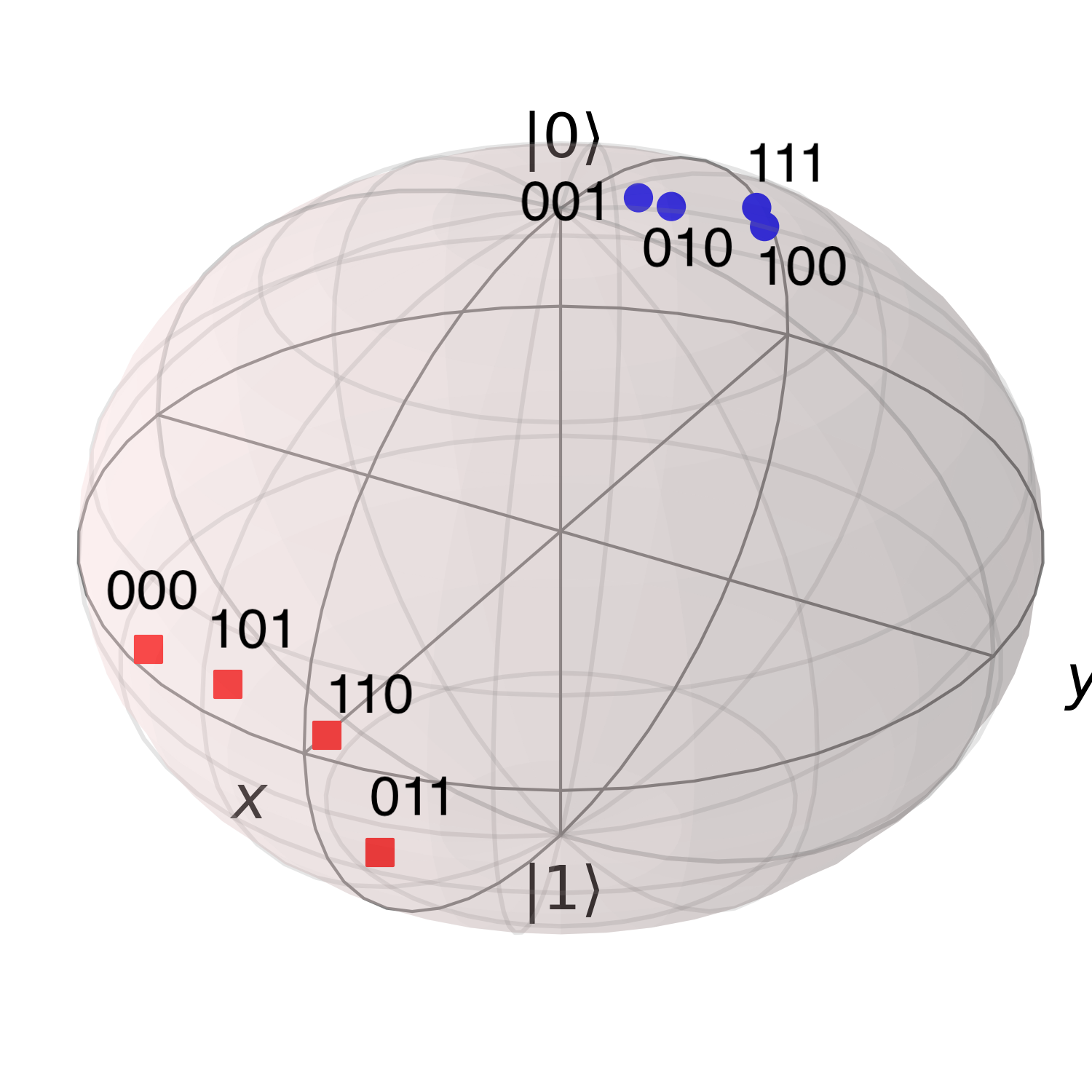}
\end{minipage}
\begin{minipage}[t]{0.3\textwidth}
\centering
    \includegraphics[width=\textwidth]{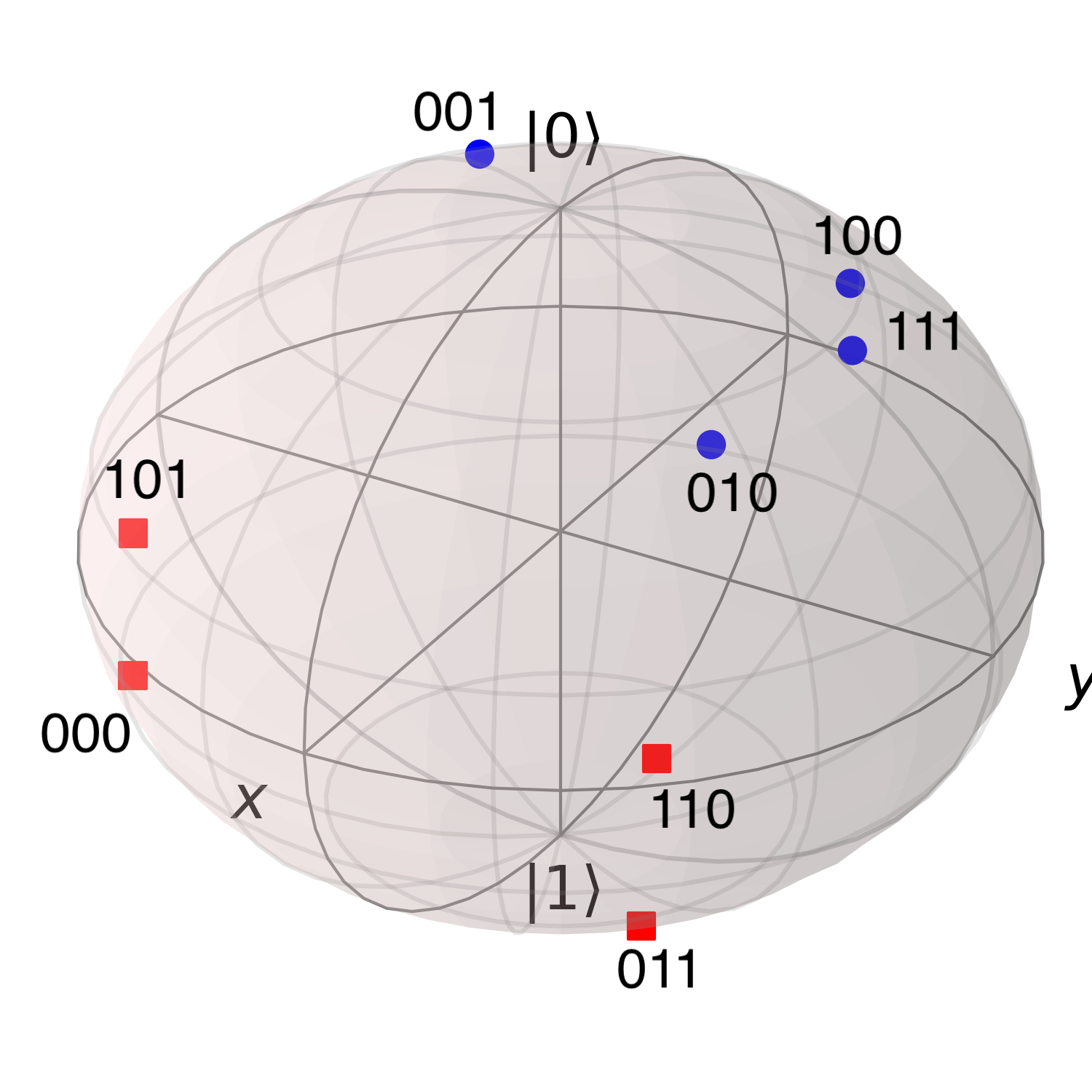}
\end{minipage}

\caption{The embeddings of $3$-bit discrete features into one-qubit quantum states drawn as colored dots on Bloch spheres whose north and south poles correspond to the classical states "0" and "1", respectively. The embedding on the left Bloch sphere is by the $(3,1)$-QRAC, while those on the middle and the right Bloch spheres are by TEs without and with regularization (as in Eq.~\ref{eq:LE_reg} with $\lambda=0.02$), respectively. The blue dots denote those with parity $1$, and the red dots parity $0$. It is clear that a $(3,1)$-QRAC is insufficient to solve the parity as there is no linear plane can separate the red from the blue dots.}
\label{fig:3bits_parity}
\end{figure*}

There are several drawbacks of mapping the above features with QRAC that fixes mapping of binary strings to quantum states. First, there are only few concrete constructions of QRACs~\cite{ANTV02,L17,IR2018}. Second, for some Boolean functions such as the parity, the fixed mapping of QRAC may result in mapping inputs to quantum states that are not ``well-separated" in the Hilbert space. This can easily be seen from mapping $8$ binary strings $\vect{x} \in \{0,1\}^3$ into one-qubit quantum states that lie in 3-dimensional Bloch sphere with QRAC as in the left subfigure of Fig.~\ref{fig:3bits_parity}; because the VC-dimension of the Boolean functions with $3$ variables is at least $4$, VQCs with QRAC can fail to find the boundary decision in the Bloch sphere. Previous work~\cite{Lloyd2020QuantumEF,Yano2020EfficientDF} suggest adding more qubits or copies of QRACs, to avoid this problem but it reduces the advantage of QRACs. We propose trainable embedding to avoid the drawbacks.  

Instead of fixing the encoded quantum spaces of the input binary strings using $(m,n)$-QRAC, we propose a method by applying a trainable circuit which is inspired by the quantum metric learning~\cite{Lloyd2020QuantumEF}. The circuit is learned with an objective to obtain maximally separating data classes in the Hilbert space. By adaptively training feature embeddings to separate the data classes, \cite{Lloyd2020QuantumEF} showed that the best measurements to obtain the classification can be realized with shallow quantum circuits and hence suitable for near-term quantum devices. 

The trainable circuits are represented as a set of unitary $\mathbf{U}(\theta_{\vect{b}})$ matrices where $\vect{b}$ is the binary string input and $\theta_{\vect{b}}$ is the set of parameters corresponding to $\vect{b}$. Namely, $(m,n)$-QRAC is replaced with a trainable quantum circuit realizing $\mathbf{U}(\theta_{\vect{b}})$. For example, the $(3,1)$-QRAC is replaced with trainable embedding using $u3(\theta,\varphi,0)$ gates for each 3-bit binary string, where $\theta$ and $\varphi$ are trainable parameters\footnote{$u3$ gate is a standard one-qubit rotational gate in Qiskit\cite{Qiskit}. The last parameter of the $u3$ gate is fixed to zero because it does not affect the transformation starting from the zero state.}. Therefore, there are 16 parameters to be trained for each qubit encoding $3$ bits. For convenience, we denote $(3,1)$-TE as the trainable embedding of $3$ bits into $1$ qubit.

However, the increasing number of parameters might possibly lead to overfitting. The center subfigure of Fig.~\ref{fig:3bits_parity} shows the result of $(3,1)$-TE for the parity of the bits. As can be shown from the left of the same figure, with QRAC the distance between any two embedded states is guaranteed to be above some threshold. On the other hand, the embedded states of the $(3,1)$-TE are clustered as shown in the middle of the same figure.

To guide the embedded states to become as diverse as those with QRACs, and so that each of them is far for the rest, we introduce a cost to penalize the entropy of the overall embedding vectors. Let us assume  $\{(\theta_i,\varphi_i)\}$ be the parameter for encoding 3-bits binary string $i$. The embedding for each $i$ is therefore $u3(\theta_i,\varphi_i,0) \ket{0} = \cos(\theta_i/2)\ket{0}+e^{\varphi_i}\sin(\theta_i/2)\ket{1}$ which corresponds to the point  $\vect{r}_i=(\sin\theta_i\cos\varphi_i, \sin\theta_i\sin\varphi_i,\cos\theta_i)$ on the $3$-dimensional Bloch sphere\footnote{See, e.g., \cite{Nielsen2000} for the correspondence of one-qubit (pure) states with points on the Bloch sphere}. From these embedding vectors $\vect{r}_1,\vect{r}_2,\ldots,\vect{r}_8$, we can compute their averages and covariance matrix, as
\begin{align*}
    \vect{\mu}&=\frac{1}{n}\sum_{i}\vect{r}_i\\
    \vect{\Sigma} &= \frac{1}{n^2}\sum_{i,j}(\vect{r}_i-\vect{\mu})(\vect{r}_j-\vect{\mu})^\top,
\end{align*}
where $n$ is the number of the embedding. In $(3,1)$-TE case, we have $n=8$ number of embedding. The determinant of the covariance matrix can be used to scale the spread factor of the embedding. By adding the negative of the determinant of the covariance matrix, i.e., $\mathcal{L}_{\text{spread}} = -\det(\vect{\Sigma})$ to the training loss to find optimal embedding parameters, we minimize the following regularized loss function: 
\begin{equation}\label{eq:LE_reg}
    \mathcal{L}_{\text{all}}=\mathcal{L} + \lambda\mathcal{L}_{\text{spread}}.
\end{equation}
The effect of regularization is shown in the right subfigure of Fig.~\ref{fig:3bits_parity}. We can see a wider spread of distribution of embedded quantum state by the regularization. A similar regularization by determinant has also been used to induce diversity of state-action vectors in reinforcement learning~\cite{OsogamiR19}.

As a comparison to classical machine learning method, TE would be similar to embedding feature that prevalently used with discrete feature such as word embedding \cite{Mikolov2013EfficientEO} to map a discrete word token into a continuous vector space. 

\paragraph{Discussion About Computational Complexity}
Assume we are applying $(3,1)$-TE instead of QRAC encoding, we want to assure that our method does not increase significant complexity when it is compared to normal VQC. In SPSA, the parameters are updated by computing symmetric finite difference gradient $f(\theta+\epsilon) - f(\theta-\epsilon)$ and update the parameter to the direction $\epsilon$. Assumed that $k$ qubits and $l$ layers of the $\vect{W}$ are used which each layer has one parameter for a qubit. Hence, in normal VQC setting, there will be $O(lk)$ parameters to update which we need to perform finite difference gradient for each parameter. If we add $(3,1)$-TE as the embedding to the model, for each qubit, there are two parameters; thus, $2k$ more parameters for each input sample. The complexity would become $O((l+2)k)$ which didn't significantly increase the complexity compared to normal VQC with QRAC.

On inference time, the complexity is exactly same as QRAC as it just need to query the two parameters $\theta$ and $\gamma$ and performs rotation on Bloch sphere. 

\section{Experiments}\label{sec:experiments}
We perform numerical experiments demonstrating the limitation of discrete feature embedding with QRAC and the possibility of trainable embedding to overcome this limitation. To this end, the embeddings are tested on parity functions, which are known as hard Boolean functions, and on real-world datasets: the Breast Cancer dataset (BC), the Titanic Survival dataset (TS), and the hand-written digits MNIST dataset. On parity functions, the experiments showed that the TE achieved higher accuracies with fewer qubits than straightforward and QRAC embeddings. We also confirmed the advantage of TE for classifying the real-world datasets. The experiments were run by Qiskit using its RyRz-variational VQCs~\footnote{The details of the RyRz variational form can be found at \url{https://qiskit.org/documentation/stubs/qiskit.aqua.components.variational_forms.RYRZ.html}}. We also confirmed the efficacy of the embeddings on TensorFlow Quantum~\cite{Broughton2020TensorFlowQA} by experimenting with QNN on the MNIST dataset.

\subsection{VQC on Parity}

The parity function is determined by the XOR of its individual input bits (or equivalently, the sum of its input bits modulo $2$). For example, for 3-bit inputs the parity function is 1 if and only if the input is in $\{001,010,100,111\}$, as colored in blue in Fig.~\ref{fig:3bits_parity}.  
We tested the embeddings on parity of bits whose lengths are $3, 6$ and $9$. 
As can be seen from the left subfigure in Fig.~\ref{fig:3bits_parity}, with the QRAC embedding the $3$-bit parity is not linearly separable. This is the main motivation to use the parity function as a test case for the embeddings.

We compared the efficacy of VQC with the embeddings by calculating the correctly classified ratio (or precision). All VQCs were optimized using SPSA~\cite{Spall92} for 400 epochs. The embeddings were as below: 
\begin{itemize}
    \item \textbf{Naive}: a qubit is used to represent a bit. %
    \item \textbf{$n\times$QRAC}: $n$ copies of $(3,1)$-QRAC encoding is used to represent the binary string. $n > 1$ is suggested to increase the chance finding a decision boundary in the Hilbert space~\cite{Yano2020EfficientDF}. 
    \item \textbf{TE}: The $(3,1)$-TE is used to encode the binary string. %
\end{itemize}

\begin{figure*}[t]
    \centering
    \includegraphics[width=0.9\textwidth]{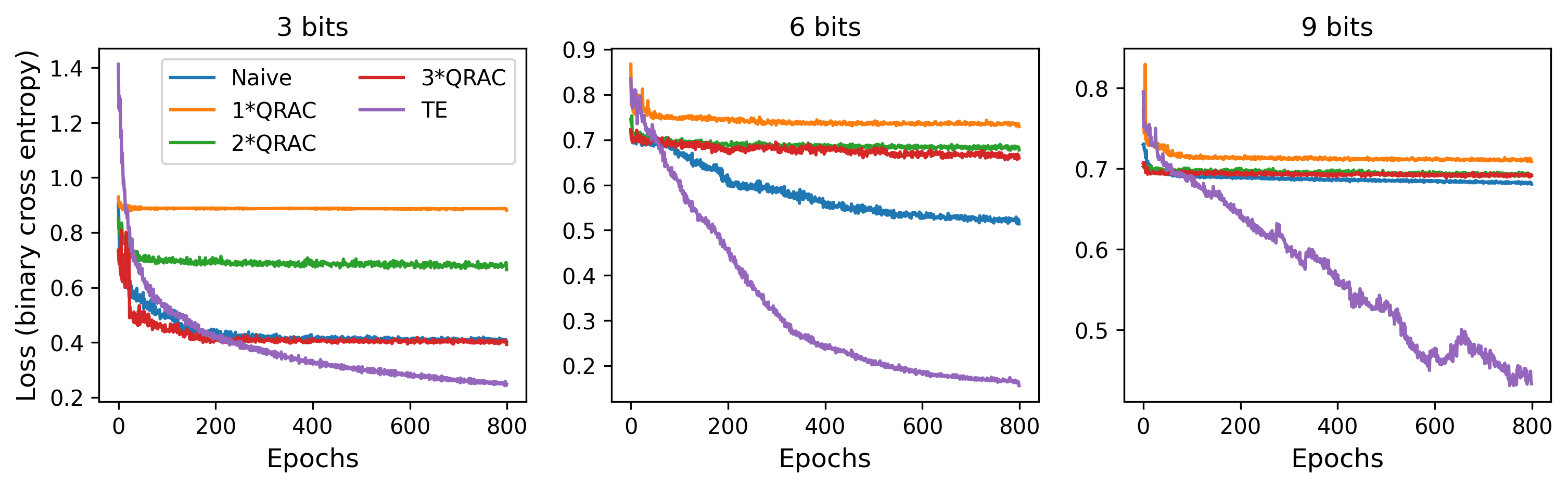}
    \caption{Plots of training losses (binary cross entropy) for learning parity of 3, 6, and 9 bits for each embedding. Note that the number of epoch in the x-axis is doubled because each epoch of SPSA consists of two stages.}
    \label{fig:PC_3bits}
\end{figure*}

Table \ref{tab:PC_main_result} shows the classified ratios of the VQCs with the embeddings. We can observe that TE achieved better precisions on the $6$-bit and $9$-bit parity despite using the fewest number of qubits. On the $3$-bit parity, its precision was only slightly less than Naive and $3$ copies of QRACs, but notice that both Naive and QRACs could not reduce the qubits. 

The parity of $6$-bit and $9$-bit showed interesting cases because TE with only $2$ and $3$ qubits, respectively, could beat Naive and QRACs that essentially store all information of the bits (by the same number of qubits). In particular, the copies of QRACs seem to only slightly increase the precision. We can further confirm the limitation of QRACs and the advantage of TE from Fig.~\ref{fig:PC_3bits} that shows the training losses of the VQCs. TEs clearly outperformed other 
embeddings, while the losses of $3$ copies of QRACs were only as good as Naive, which is expected. This is because in that case one can extract all values of bits from the qubits.

\begin{table}[!t]
\caption{Classified ratios of VQCs with various discrete-feature embeddings on parity functions.}
\label{tab:PC_main_result}
\hbox to\hsize{\hfil
\begin{tabular}{l|c|r}
\hline
\multicolumn{3}{c}{\textbf{3-bit parity} (8 strings)} \\
\hline\hline
Method & \#qubits & Classified ratio \\\hline
Naive & 3 & \textbf{1.000}\\
$1\times$QRAC & 1 & 0.500 \\
$2\times$QRAC & 2 & 0.750\\
$3\times$QRAC & 3 & \textbf{1.000}\\
TE & 1 & 0.925\\\hline
\multicolumn{3}{c}{\textbf{6-bit parity} (64 strings)} \\
\hline\hline
Naive & 6 & \textbf{1.0}\\
$1\times$QRAC & 2 & 0.625 \\
$2\times$QRAC & 4 & 0.453 \\
$3\times$QRAC & 6 & 0.625 \\
TE & 2 & 0.884  \\\hline
\multicolumn{3}{c}{\textbf{9-bit parity} (512 strings)} \\
\hline\hline
Naive & 9 & 0.559 \\
$1\times$QRAC & 3 & 0.494 \\
$2\times$QRAC & 6 & 0.517 \\
$3\times$QRAC & 9 & 0.525 \\
TE & 3 & \textbf{0.786} \\\hline
\end{tabular}\hfil}

\end{table}

\subsection{VQC on Breast Cancer dataset (BC)}
The BC dataset contains 286 samples, each of which consists of $9$ discrete features, along with a cancer recurrence as a target variable (201 samples of no-recurrence, 85 samples of recurrence). 
We selected $4$ most important features determined by a random forest algorithm \cite{10.1023/A:1010933404324}, which are \textit{tumor-size}, \textit{breast-quad}, \textit{deg-malig}, and \textit{age}. The features have 11, 6, 3, and 6 different categorical values respectively.

Due to imbalance in the labels of the dataset, we doubled the positively-labeled data in the training set to encourage true positive predictions, and evaluated F1 score in addition to accuracy (precision). We tested the VQCs with the RyRz variational form, all of which were optmized using SPSA for 200 epochs, using the three different embeddings as follows.

\begin{itemize}
    \item \textbf{ZZ}: the baseline mapping for real-valued features as used in~\cite{Havlcek2019SupervisedLW}. Each feature is simply encoded into an ordinal value, and then embedded using the ZZ feature map (ZZ).
    
    \item \textbf{QRAC}: the mapping used in~\cite{Yano2020EfficientDF}. Each feature is encoded into an ordinal value, and then converted into a bit string. The concatenated bit string is encoded with $(3,1)$-QRACs.
    
    \item \textbf{TE}: similar to QRAC up to the concatenated bit string.  $(3,1)$-TE is used to embed the concatenated string. Reg.TE denotes the a regularized model as in Eq.~\ref{eq:LE_reg}. We obtained the hyperparameter by searching on $\lambda$ on the log-scale $[10^{-5}, 10^2]$ and chose $\lambda = 0.02$ to prevent overfitting. (In general, $\lambda\sim 10^{-2}$ likely to yield best result).
\end{itemize}

The experimental results shown in Table~\ref{tab:BC_main_result} suggest that feature maps with QRAC and TEs significantly outperformed ZZ feature map (ZZ), with TEs had better accuracies than QRAC. Without regularization, however, TE showed slight overfitting to the training data. TE with regularization (Reg.TE) avoided the overfitting. 

\begin{table}[!t]
\caption{Accuracies and F1 scores of 4-fold cross validated VQCs on the BC dataset, with four different discrete feature mappings. All mappings require four qubits.}
\label{tab:BC_main_result}
\hbox to\hsize{\hfil
\begin{tabular}{l|ll}\hline\hline
Method & Train acc./F1 & Test acc./F1 \\\hline
ZZ & 0.606/0.475 & 0.549/0.419 \\
QRAC & 0.707/\textbf{0.541} & 0.682/0.483  \\
TE & \textbf{0.722}/0.534 & 0.681/0.452 \\
Reg.TE & 0.714/0.534 & \textbf{0.702}/\textbf{0.511} \\\hline
\end{tabular}\hfil}
\end{table}

\subsection{VQC on Titanic Survival dataset (TS)}
The TS dataset consists of 11 features, including both discrete and continuous ones. 
The target variable is the survival of the passengers. 
This dataset is split into training and test sets whose size are 891 and 418 samples respectively.

We selected 4 most important features as ranked again by the random forest algorithm. The features are \textit{sex}, \textit{age}, \textit{pclass}, \textit{fare} which are mixtures of continuous and discrete ones. 
We performed two types of experiments: VQCs with embedding both continuous (with re-scaling) and discrete features, as well as VQCs with discrete features (after by ordinal representation of the continuous ones). 

As for the mapping of the discrete features in both experiments, we implemented four distinct feature mappings as in the BC dataset, where the two rescaled continuous features in the first experiment were embedded with the ZZ feature map. The RyRz-variational form of the VQCs were of depth $4$ and were optimized using SPSA for 300 epochs in all experiments.

Table~\ref{tab:TS_main_result} shows the results of the two experiments. The results indicate that encoding continuous features as ordinal values slightly improves the overall performances of the models. However, F1 scores for the mixed TE(dis.) + ZZ(cont.) models were higher than the discrete-only TE(dis.) counterparts. The results also confirm that QRAC and TE were better than the baseline ZZ feature maps, and TE with regularization fairly to be on a par with QRAC.

\begin{table}[tb]
\caption{Accuracies and F1 scores of 4-fold cross-validated VQCs on the TS dataset embedding two ordinal features and two rescaled continuous features (top four rows), and those embedding ordinal-encoded features (bottom four rows).For both experiments, the top ZZ methods used 4 qubits, while QRACs and TEs used 3 qubits.}
\label{tab:TS_main_result}
\hbox to\hsize{\hfil
\begin{tabular}{l|cc}\hline\hline
Method & Train acc./F1 & Test acc./F1 \\\hline
ZZ(dis.+cont.)  & 0.729/0.636 & 0.706/0.610  \\
QRAC(dis.)+ZZ(cont.)& \textbf{0.777}/0.699 & \textbf{0.767}/0.687 \\
TE(dis.)+ZZ(cont.) & 0.749/0.711 & 0.739/0.706 \\
Reg.TE(dis.)+ZZ(cont.)  & 0.764/\textbf{0.723} & 0.756/\textbf{0.713}\\\hline
ZZ(dis.) & 0.740/0.643 & 0.711/0.601 \\
QRAC(dis.)  & 0.785/\textbf{0.721} & 0.772/\textbf{0.707}  \\
TE(dis.) & 0.771/0.704 & 0.761/0.692 \\
Reg.TE(dis.) & \textbf{0.788}/0.714 & \textbf{0.776}/0.702\\\hline
\end{tabular}\hfil}

\end{table}

\subsection{QNN on hand-written digit MNIST}

MNIST is one of the most well-known datasets that consists of black-white hand-written digit images of size $28\times 28$. Such instances of images can be fed easily into classical machine learning. However, due to the limited number of qubits of near-term quantum devices, the size of those images must be reduced by some preprocessing in order to be usable in quantum machine learning. We followed the preprocessing shown at the tutorial of TensorFlow Quantum\footnote{As detailed in \url{https://www.tensorflow.org/quantum/tutorials/mnist}}, which is similar to the preprocessing in~\cite{Farhi2018ClassificationWQ} . Only images of digits 3 and 6 were selected while those labelled both 3 and 6 were omitted. The selected images were resized down to $4\times4$, and their pixels were rounded to 0s and 1s to obtain binary string representations. The classification task is binary decision (i.e., either 3 or 6). We compared various embeddings of the images as the following. 

\begin{itemize}
    \item \textbf{Naive 16px}: a qubit for a pixel/bit of the $4\times 4$ pixels. This was also used in~\cite{Farhi2018ClassificationWQ}. 
    \item \textbf{Naive 8px}: a qubit for a pixel of those at the second and third rows of the $4\times 4$ pixels. The purpose of this method is to observe the performance loss/gain if we reduce the number of qubits by ignoring parts of the inputs.
    \item \textbf{QRAC/TE}: the flatten $4\times 4$ pixels, and thus 16 bits were embedded with $(3,1)$-QRAC or $(3,1)$-TE. Thus, in total 6 qubits were used.
    \item \textbf{Conv QRAC/TE}: convolutional embedding of the $4\times 4$ pixels. C-QRAC and C-TE were used for every 3 consecutive pixels of each row and column with stride 1. Formally, suppose that $\vect{b}$ is the $4\times 4$ input image. Every 3 pixels at $i$-th row: $b_{ij}b_{i(j+1)}b_{i(j+2)}$ for $1\leq i\leq 4$ and $1\leq j\leq 2$, and 3 pixels at $j$-th column: $b_{ij}b_{(i+1)j}b_{(i+2)j}$ for $1\leq i\leq 2$ and $1\leq j\leq 4$ were encoded with $(3,1)$-QRAC or $(3,1)$-TE, resp. Therefore, as in \textit{Naive 16px}, in total 16 qubits were used. 
    
    \item \textbf{Conv (4,1)-TE}: convolutional embedding on the $4\times 4$ image with kernel size $2\times 2$ and stride 1. Each $2\times 2$ block was embedded using $(4,1)$-TE. Formally, suppose that $\vect{b}$ is the $4\times 4$ input image, we embedded every $2\times2$ pixels $b_{ij}b_{i(j+1)}b_{(i+1)(j)}b_{(i+1)(j+1)}$ for $1\leq i,j\leq 3$ with $(4,1)$-TE. In total, 9 qubits were used for the embedding. Note that $(4,1)$-QRAC is impossible~\cite{HINRY06}.  
\end{itemize}

\begin{table}[tb]
\caption{Experimental results of QNNs on MNIST dataset. Training and testing accuracies of various discrete feature embeddings with fewer qubits than Naive embeddings of~\cite{Farhi2018ClassificationWQ} are shown.}
\label{tab:MNIST_main_result}
\hbox to\hsize{\hfil
\begin{tabular}{l|r|cc}\hline\hline
Method & \#qubits & Train acc. & Test acc. \\\hline
Naive 8px & 8 & 0.872 & 0.891 \\
QRAC & 6 & 0.883 &	0.898 \\
TE & 6 & \textbf{0.903}	& \textbf{0.912}  \\\hline
Naive 16px & 16 & 0.902 & 0.908  \\
Conv QRAC & 16 & 0.896 & 0.898 \\
Conv TE & 16 & 0.911 & 0.914 \\
Conv $(4,1)$-TE & 9  & \textbf{0.911} & \textbf{0.917}\\
\hline
\end{tabular}\hfil}

\end{table}

\begin{figure}[tb]%
\centering
\includegraphics[width=0.45\textwidth]{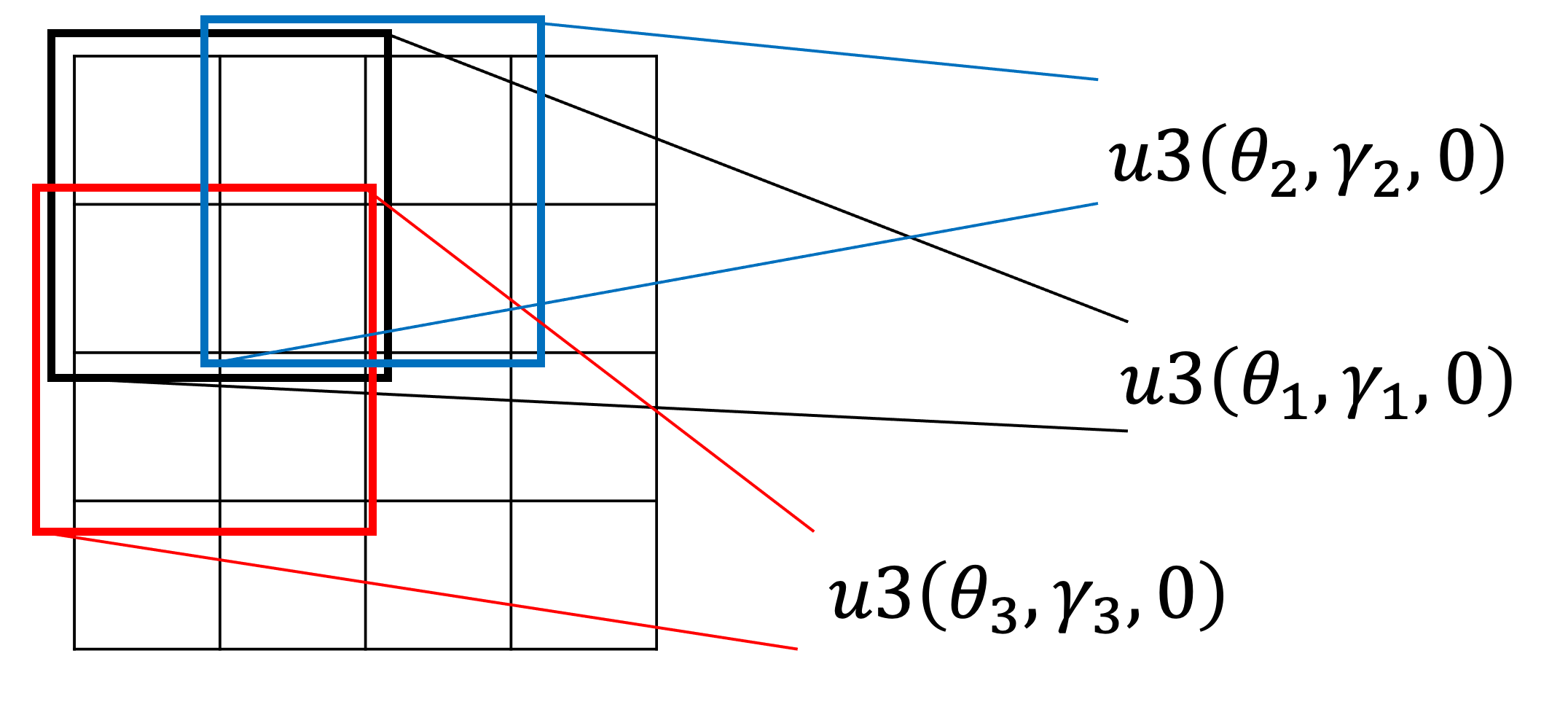}

\caption{Illustration of convolutional embedding by (4,1)-TE on $4\times4$ MNIST image.}
\label{fig:MNIST_41}
\end{figure}

For every QNN with a particular embedding, we employed the same structure of circuit $\mathbf{W}(\vect{\theta})$ that consisted of mainly two layers, the first layer consisted of Ising $XX$ gates controlling from every input qubits toward the output qubit. The second layer consisted of Ising $ZZ$ gates controlling from every input qubits as the first layer\footnote{The Ising $XX$ and $ZZ$ gates are also used in \url{https://www.tensorflow.org/quantum/tutorials/mnist}}. The binary cross entropy was used for training loss. We performed 5 different runs for each embedding method using Adam \cite{Kingma2015AdamAM} as an optimizer for 10 epochs and reported their average accuracy. The experimental results are as in the Table \ref{tab:MNIST_main_result}.

As can be seen from the table, by using QRAC or TE we can embed the images with fewer qubits. Moreover, the accuracies of the embedding were better compared to the Naive method with 8 qubits, and were comparable to that with 16 qubits. In particular, we can see that Convolutional TEs performed better than others, even for Convolutonal $(4,1)$-TE with only 9 qubits. The drawback of the Convolutional TE is the significant increase of the parameters for the embeddings, but not all combination of 4-bit binary strings appear; hence, some parameters can be ignored.

\subsection{TE on Real Quantum Devices}

\begin{table}[tb]
\caption{Experimental on real device for 6-bits parity and TS dataset using TEs. 6-bits parity result is from a single run. TS result is the average of 4 folds accuracy.}
\label{tab:real_device}
\hbox to\hsize{\hfil
\begin{tabular}{l|r|cc}\hline\hline
Dataset & \#qubits & Train acc. & Test acc. \\\hline
6-bit parity & 2 & 1.000 & - \\
TS & 3 & 0.715 & 0.709 \\

\hline
\end{tabular}\hfil}

\end{table}

\begin{figure}[tb]
    \centering
    \includegraphics[width=0.45\textwidth]{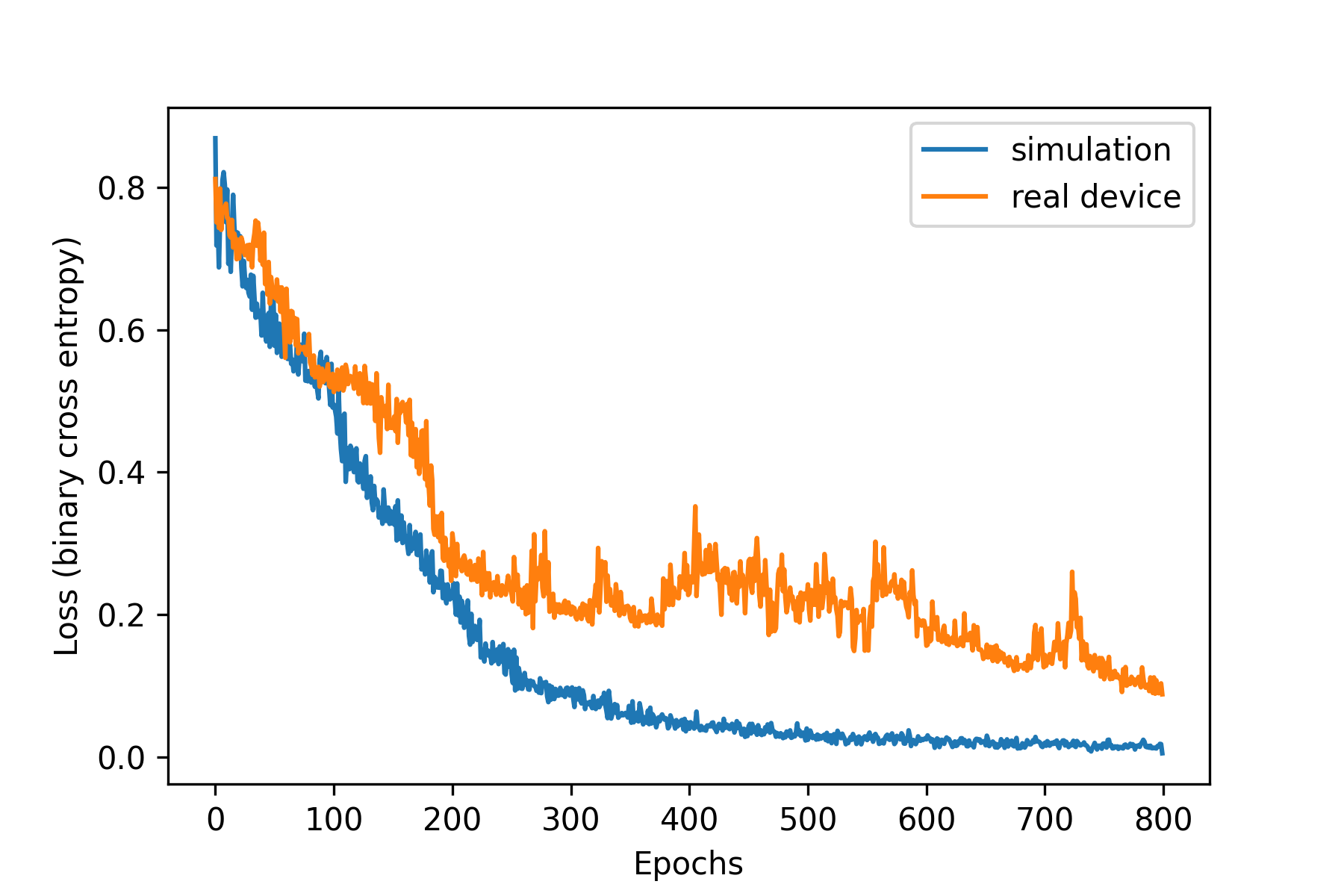}
    \caption{The training losses on 6-bit parity comparing simulation and real device.}
    \label{fig:6-bit_real_vs_sim}
\end{figure}

\begin{figure}[tb]%
\centering
\includegraphics[width=0.45\textwidth]{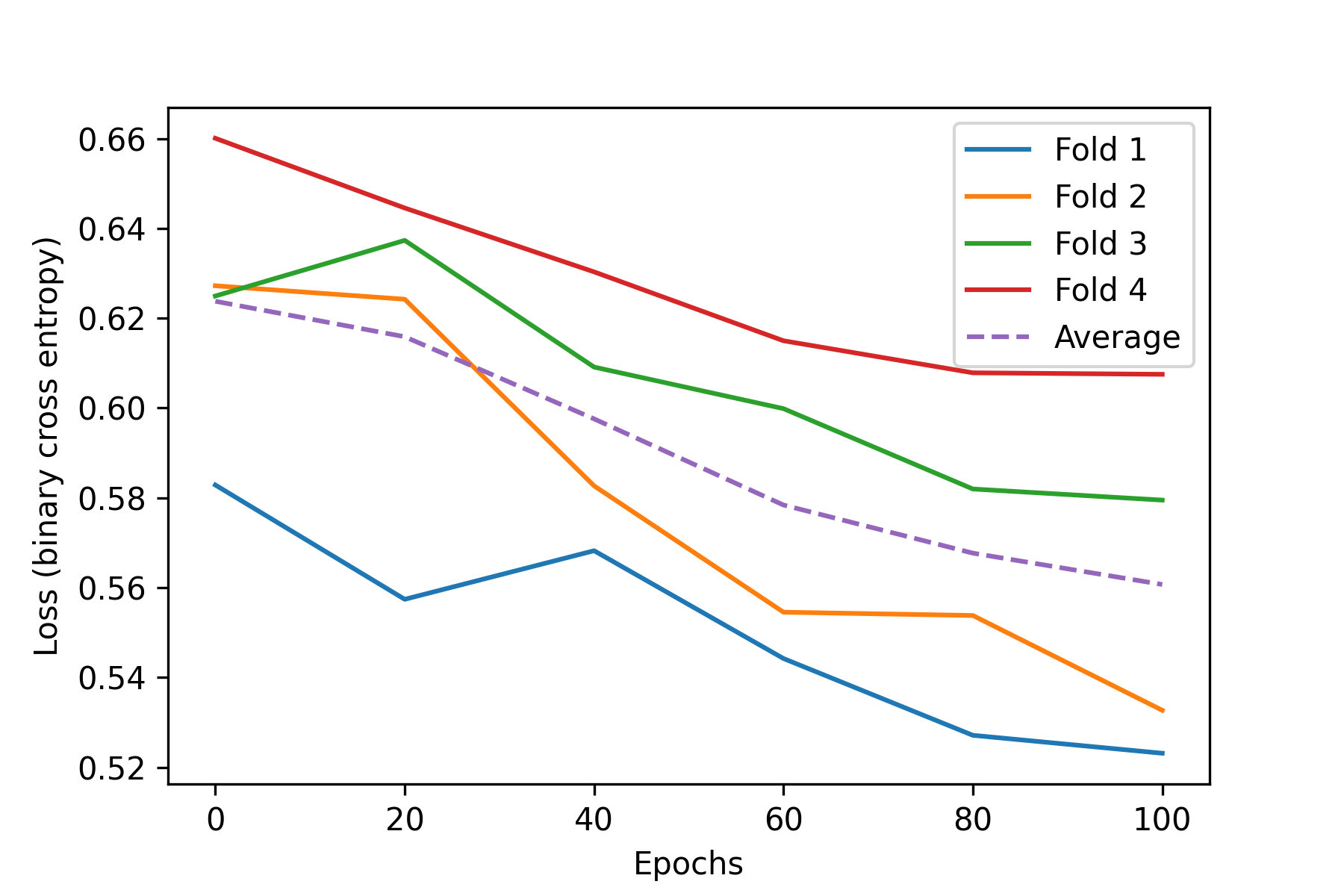}

\caption{The training losses on TS for each fold on real device. Losses are recorded every 20 epochs.}
\label{fig:TS_loss_real_device}
\end{figure}

We further shows that our algorithm can be run on the real device. From June 1 to June 5, 2021 we experimented with the \textsf{ibmq\_athens}, a real-device backend of IBM Quantum System, using the \textit{fairshare} run mode to classify the 6-bit parity and TS with discrete features. We conducted experiments on the 6-bit parity with 400 epochs for a single run on the first two qubits of the device. We conducted experiments on the TS by splitting the dataset into four folds and learnt for each fold with 100 epochs of TE regularized with $\lambda=0.02$. As the TS needs 3 qubits for its experiment and the connectivity of \textsf{ibmq\_athens} is linear, to reduce the number of CNOT gates we used a linearly entangled RyRz on the 0-th, 1-st, and 2-nd qubit of the device instead of fully entangled one used previously to obtain Table~\ref{tab:TS_main_result}. 

Experimental results show that the the trainable embedding could be trained on real quantum devices to obtain accuracies as shown in Table \ref{tab:real_device}. We choose the seed with best result from 6-bit parity experiment on simulation and re-run it on real device. Under the same random seed, TE could fully classify, attaining accuracy of 1.0, the 6-bit parity problem on both simulation and real device. We visualize the training loss in Figure~\ref{fig:6-bit_real_vs_sim}, showing that both simulation and real device converge. However, we can observe that there are higher fluctuation in real device case. We speculate that this might be the result from noises in real device.

Due to time limitation we trained the TS for only 100 epochs instead of 300 epochs to obtain the previous Table~\ref{tab:TS_main_result}. Despite the limitation, the experimental results showed that TE could learn well as it improved to get less training loss through epochs as shown in Fig.~\ref{fig:TS_loss_real_device}. Even with significantly less number of epochs and less entanglement power due to device's connectivity, we obtained accuracy well above 70\% and we could expect to obtain better with more epochs and entanglement. We concluded that TE can be trained effectively to classify a dataset with discrete features on real quantum devices. 

\section{Conclusion}\label{sec:conclusion}
We propose trainable discrete feature embeddings by combining QRAC with the quantum metric learning. Our trainable embeddings show a remarkable advantage on the parity over embedding with QRACs~\cite{Yano2020EfficientDF} which needs additional qubits to overcome the linear separability limitation due to their fixed encoding. Furthermore, trainable embeddings also provide good classification performance with fewer qubits on various real-world datasets as demonstrated numerically through the framework of VQCs and QNNs. Another advantage of trainable embeddings is that it is not limited by the existence of QRACs, i.e., we show that $(4,1)$-TE is possible to train and be used for classification achieving better or comparable accuracies with fewer qubits despite the non-existence of $(4,1)$-QRAC~\cite{HINRY06}. For this reason, we believe our trainable embedding is flexible and can provide satisfying results in various tasks of quantum machine learning. In addition, we show that our algorithm can be run on current real device even more parameters are introduced. We hope that this work can open more possibility to quantum machine learning on other fields such as Natural Language Processing (NLP) that mostly deals with discrete features.

Lastly, there are clearly many directions for future work, including to develop trainable embedding with more qubits, to find good heuristics for the initialization as it heavily affects the final result of embeddings, and to design generalized $(m,n)$-TEs with regularization that provides theoretical guarantees. 

\section{Acknowledgement}
We would like to thank Prof. Hiroshi Imai of University of Tokyo for his support in the lectures of Introduction to Quantum Computing from April to July, 2020 during which parts of this paper were conceived. RR would also like to thank the TAs of the lectures, Hyungseok Chang and Atsuya Hasegawa, for their help, and Yutaka Shikano of Keio University for valuable feedback and suggestion. 

\bibliography{ref.bib}{}
\bibliographystyle{IEEEtran}

\end{document}